\documentclass[aps,pre, twocolumn, groupedaddress]{revtex4-1}

\usepackage{graphicx}
\usepackage{dcolumn}
\usepackage{bm}
\usepackage{color}

\begin{document}


\title{Collective dynamics of self-propelled particles with variable speed}


\author{Shradha Mishra}
\author{Kolbj{\o}rn Tunstr{\o}m}
\author{Iain D.~Couzin}
\affiliation{Department of Ecology and Evolutionary Biology, Princeton University, Princeton, NJ 08544}

\author{Cristi\'an Huepe}
\affiliation{614 N Paulina St, Chicago IL 60622, USA}


\date{\today}

\begin{abstract}
\noindent Understanding the organization of collective motion in biological systems is an ongoing challenge. In this Paper we consider a minimal model of self-propelled particles with variable speed. Inspired by experimental data from schooling fish, we introduce a power-law dependency of the speed of each particle on the degree of polarization order in its neighborhood. We derive analytically a coarse-grained continuous approximation for this model and find that, while the variable speed rule does not change the details of the ordering transition leading to collective motion, it induces an inverse power-law correlation between the speed or the local polarization order and the local density. Using numerical simulations, we verify the range of validity of this continuous description and explore regimes beyond it. We discover, in disordered states close to the transition, a phase-segregated regime where most particles cluster into almost static groups surrounded by isolated high-speed particles. We argue that the mechanism responsible for this regime could be present in a wide range of collective motion dynamics.
\end{abstract}
\maketitle



\section{Introduction}

Bacterial colonies~\cite{benjacob}, insect swarms~\cite{rauch}, bird flocks~\cite{starling}, and fish schools~\cite{hubbard} are all examples of biological systems that display distinct collective motion. While they differ in many specific aspects, they also share an important characteristic: Each one is made up of individual organisms that are self-propelling and move guided by interactions with their surrounding neighbors. When we strip down collectively moving systems in this way, we obviously leave out important biological details, but in return we obtain a good starting point for a theoretical understanding of collective motion. Our central concern in this paper is to introduce and study a very simple model (a {\it minimal} model) where individuals move with variable speed, and to show through theory and simulations that it displays new properties not found in models with constant speed. 

In recent years, research on self-propelled (or {\it active}) systems has grown steadily. Starting with the seminal work by Vicsek et al.~\cite{vicsek}, we now find a range of theoretical, numerical and experimental studies \cite{tonertu,tonertusr,sriramaditi,chuepe1,chateall,aparnamarchetti,shradhasr,iaincouzin1,kudrolli,vijayn} concerning a number of aspects of their dynamics. 
In Vicsek's original work~ \cite{vicsek}, a minimal model of self-propelled collective motion---now known as the {\it Vicsek model}---was introduced, similar in flavor to the equilibrium XY-model \cite{xy}, but exhibiting new physical properties due to its  non-equilibrium nature.

In the standard Vicsek model, each individual is described as a point particle with a given position and heading direction. All particles advance with the same constant speed and decide their next heading direction based on the headings of all particles in their local neighborhood. Particles are also subject to noise, which alters their chosen direction. Under these simple conditions, the Vicsek model exhibits a dynamical phase transition from a disordered state (with no polarization order) to an ordered state (where particles align and advance together) as the noise level is decreased, or the density of particles increased.

Following its initial formulation, many variations of the original Vicsek model \cite{chateall,vicsekall} have been introduced to carry out analytical \cite{tonertu,aparnamarchetti,shradhasr, shradhaapranacristina} and numerical \cite{chatepre,vicsek} studies of collective motion. However, almost all of these models consider particles that advance with equal constant speed. But in biological systems the speed of each individual could vary in response to its neighbors' dynamics, and it is reasonable to expect this variation to be important for the resulting collective motion.
Furthermore, in recent experiments on fish schooling, it has been observed that fish swimming in disordered regions typically move slower than those in ordered ones \cite{iainexp,PNASYael}. This observation supports the idea that variable speed may play an important role in the collective dynamics of real groups of self-propelled individuals.

In this Paper, we introduce an experimentally motivated minimal model describing the collective motion in groups of self-propelled particles with variable speed, and study it analytically and numerically.
The model is almost identical to the standard Vicsek model, but here \textit{both} the speed and orientation of each particle depends on the state of its local neighborhood. While the choice of speed dynamics is motivated by our experimental results on schooling fish, we emphasize that the model is not intended to function as a detailed replicate of this biological system. Note that models similar to the one presented here were used in two recent numerical studies focusing on how variable speed can enhance convergence to an ordered state \cite{preadaptive}. These studies, however, did not address the  analytical results or novel spatial dynamics discussed here.
Starting from our model, we derive the corresponding hydrodynamic equations of motion for the density of particles and for the order parameter field, which undergoes a symmetry breaking transition at the critical point, like in the original Vicsek model. Using these equations, we find an analytical relation between the coarse-grained particle speed and the local density of particles, which we verify numerically.
We then study our variable speed model numerically, characterizing its ordered and disordered phases as a function of the noise, mean particle density, and variable speed rule. In particular, we discover in the disordered regime a novel state where static cluster are formed, containing particles moving at speeds close to zero. Within these clusters, each particle receives conflicting heading information from its neighbors and thus advances at low speeds, which in turn limits its ability to spread its own heading information throughout the group. We hypothesize that this mechanism could also be present in more realistic systems.

The Paper is organized as follows. In Section II we introduce our variable speed model using new experimental results to motivate the particle speed dynamics. In Section III we derive the corresponding coarse-grained hydrodynamic equations of motion and compare them to the constant speed case. Section IV presents numerical results that confirm the validity of the analytical approximations and explore regimes beyond these, where the hydrodynamic description is no longer valid. We study here the static clusters described above. Finally, Sections V and VI  contain our discussion and conclusions. 




\section{variable speed model and experimental motivation \label{Model}} 

\subsection{Experimental background}

\begin{figure}
\begin{center}
\includegraphics[width=8.6cm]{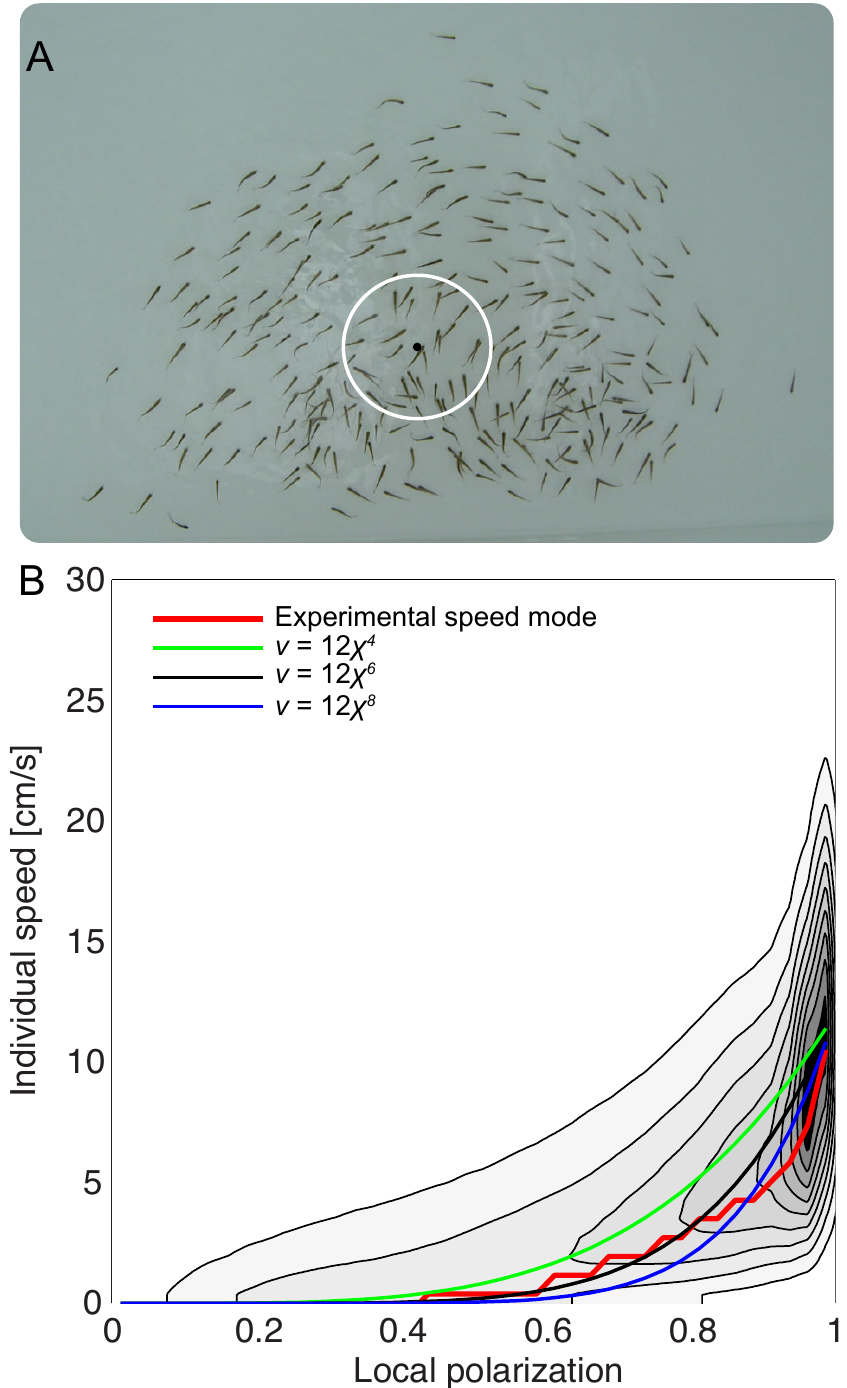}
\caption{
(A): Snapshot of experimental system with 300 golden shiners swimming in a shallow tank. The white circle defines the local neighborhood $U_i$ of the focal fish $i$ marked by the black dot. The full experiment consists of $3 \times 56$ minutes of video at 30 fps. 
(B): Experimental relationship between the local polarization order $\chi$ around an individual fish and its speed $v$. 
The contour lines delineate a 2d surface obtained by measuring the probability distribution of $v$ for every given value of $\chi$. 
The overlaid curves display the mode of these distributions, together with the $v = v_M {\chi}^{\gamma}$ relationship used in our variable speed rule, for different values of $\gamma$. We set here $v_M = 12$ in order to make $v$ coincide at $\chi=1$ with the mode of the corresponding experimental distribution.
}
\label{fig01}
\end{center}
\end{figure}

Fish schools are a clear example of systems exhibiting collective motion. To motivate the choice of speed dynamics in our variable speed model we examined experimentally the relationship between individual speed and local polarization order in a school of 300 golden shiners (\textit{Notemigonus crysoleucas}) swimming freely in a shallow tank. A snapshot from the experiment is shown in Fig.~\ref{fig01}(a). The state of each individual $i$ at a given time $t$ is defined by its position $\vec{r}_i(t)$ (defined as the centroid of the fish's image), speed $v_i(t)$ and heading direction unit vector $\hat{n}_{i}(t)$, determined from automated video tracking. Further details on the experimental setting, protocol and tracking system can be found in Katz et al.~\cite{PNASYael}.

To quantify the level of heading alignment in the local neighborhood of an individual fish $i$ we first define the set $U_i$ that contains all neighbors $j$ within a given interaction radius $\tilde{r}$, such that $|\vec{r}_i - \vec{r}_j| \le \tilde{r}$. Note that $U_i$ also includes the focal fish $i$. We then define $\chi_{i}$ as a measure of the local polarization order around individual $i$, given by
\begin{equation}
\chi_{i}(t) = \frac{1}{\mathcal{N}_{i}} \left| \sum_{ j \in U_{i} }{\hat{n}}_{j}(t) \right|,
\label{chi}
\end{equation}
where $\mathcal{N}_i$ is the number of individuals in $U_i$.

We analyzed the experimental data by finding for every individual fish in each of the 300 000 frames recorded its speed $v_i(t)$ and local order $\chi_i$. We use here $\tilde{r} \approx 15.5$ cm, but verified that similar results are obtained for different values of $\tilde{r}$. This interaction zone is illustrated in Fig.~\ref{fig01}(a) by a white circle centered around a focal fish marked by a black dot. We then generated a histogram of the $v_i(t)$, $\chi_i$ data. For each bin value of $\chi$, we normalized the speed in order to get the probability distributions for the speed as a function of the local polarization order. The resulting data is visualized as a smoothed contour plot in Fig.~\ref{fig01}(b).
It is apparent that there is typically a strong (superlinear) relationship between the speed of an individual fish and the alignment of its local neighborhood. 
Note that this result does not reveal anything about the causality, since we only show here correlations within the data. These could be a consequence, for example, of having individuals that advance slower when they find themselves in a disordered environment or that are unable to align to slow-moving neighbors.  

\subsection{variable speed model}

Our model system consists of $N$ polar particles moving in a plane with periodic boundaries.  At every time-step, each particle updates its position according to the rule 
\begin{equation}
\vec{r}_{i}(t + \Delta t) = \vec{r}_{i}(t) + v_{i}(t) \hat{n}_{i}(t).
\label{position}
\end{equation}
To compute the new position we must first define update rules for the speed and direction. The latter is given by 
\begin{equation}
\hat{n}_{i}(t+1) = \frac{1}{W_{i}} \left[ \sum_{j \in U_{i}} \hat{n}_{j}(t) + \mathcal{N}_i {\bf \eta}_{i} \right],
\label{velocity}
\end{equation}
were $W_i$ is a normalization factor chosen so that $\hat{n}_i$ is a unit vector. Noise is introduced by adding the randomly oriented vector $\vec{\eta}_i = \eta \left( \cos \phi_{i}, \sin \phi_{i} \right)$, with $\eta$ the noise intensity and $\phi_{i}$ a uniformly distributed random variable in the interval $[-\pi, +\pi]$.

Guided by the experimental results, we formulate a minimal variable speed model by considering a simple power-law relationship between the local order $\chi_i$ around particle $i$ and the particle speed $v_i$. We write it as
\begin{equation}
v_{i}(t) = v_M \left[ \chi_{i}(t) \right]^{\gamma}.
\label{speed}
\end{equation}
If all neighbors within the interaction range $\tilde r$ of particle $i$ have a similar heading, we will have $\chi_{i} \simeq 1$ and the particle speed will be close to its maximal value $v_M$. By contrast, in disordered regions $\chi_{i} \simeq 0$ and particles will advance with speed close to zero. Note that for any $\gamma$ an isolated particle will move with maximal speed $v_M$, since a particle is always contained within its own neighborhood, $i \in U_i$, which implies $\chi_{i} = 1$ in this case. 
The exponent $\gamma$ controls the shape of the curve that relates local order and speed, as shown in Fig.~\ref{fig01}(b). For $\gamma = 0$, we recover the fixed speed model. Other examples of the variable speed rule in Eq.~(\ref{speed}) for different values of $\gamma$ are overlaid on Fig.~\ref{fig01}(b).
We observe that the variable speed rule greatly simplifies the individual dynamics in that it replaces the speed distribution expected for a given $\chi_{i}$ by a single imposed speed value. It captures, however, the qualitative  dependency of the typical (most common) value of $v_{i}$ on the local polarization order. If we would consider the mean of the speed distribution, instead of its mode, a similar curve would be obtained but now $v_{i}$ would not approach zero for low $\chi_{i}$ values.
Either way, our minimal variable speed rule (\ref{speed}) was not formulated to capture these subtleties and provides a reasonable simple qualitative approximation of the actual speed dynamics. 

Keeping $N$ and $\tilde r$ fixed, the main parameters of our model are the mean number density of particles (per interaction zone area) $\rho_s = {N \pi \tilde r^2} / {L^2}$, the noise intensity $\eta$, and the variable speed exponent $\gamma$. In our numerical study, we will explore different $\rho_s$ regimes by varying the system size $L$ and study the order-disorder transition as a function of $\eta$ and $\rho_s$.



\section{Analytical results \label{Analytical}}

Systems of active particles are by construction out of equilibrium, since energy is being continuously injected at the particle level for self-propulsion. Therefore, free energy functional methods cannot be applied. It is, however, possible to study the coarse-grained dynamics of self-propelled particles in terms of a set of hydrodynamic equations, which are derived either using symmetry arguments~\cite{tonertu,sriramaditi} or directly from the underlying microscopic model~\cite{shradhasr,aparnamarchetti,bertin}. We choose here the latter approach to obtain hydrodynamic equations for the variable speed model. Specifically, we will find equations for the coarse-grained local density and polarization order parameter fields and compare them to the constant speed case \cite{shradhasr}.
The derivation is carried out in the spirit of the Ginzburg-Landau theory \cite{chaikinlubensky}, that is, in an approximation that assumes slow modulations in time and space of the local order parameters considered.
We follow here the same approach detailed in \cite{shradhasr}, but applying it to a variable speed case.

We begin by defining the coarse-grained local density field as
\begin{equation}
\rho(\vec{r}, t) = \sum_{i=1}^N\delta(\vec{r}-\vec{r}_{i}),
\label{defrho}
\end{equation}
the polarization order parameter field as the vector
\begin{equation}
\vec{P}(\vec{r}, t) = \frac{\sum_{i=1}^N {\hat{n}_{i}(t) \, \delta(\vec{r}-\vec{r}_{i})}}{\rho(\vec{r}, t)},
\label{defp}
\end{equation}
and the traceless symmetric tensor, 
\begin{equation}
{\bf Q}(\vec{r}, t) = \frac{\sum_{i=1}^N 
{ \left[ \hat{n}_{i}(t) \, \hat{n}_{i}(t) \, - \frac{1}{2} {\bf 1} \right] \delta(\vec{r}-\vec{r}_{i})}}{\rho(\vec{r}, t)},
\label{defq}
\end{equation}
which would correspond to the nematic order parameter field in systems with apolar order \cite{shradhasr}.
Here, $\delta( \cdot )$ is the Dirac delta function, ${\bf 1}$ is the identity matrix, and the outer product
$\hat{n}_i \hat{n}_i$ yields a $2 \times 2$ matrix with entries ${n}_i^{\alpha} {n}_i^{\beta}$, where $n_i^\alpha$ 
and $n_i^\beta$ are the $\alpha$ and $\beta$ components of unit vector $\hat{n}_i$, respectively.
Using the update rules in Eqs.~(\ref{position}), (\ref{velocity}) and  (\ref{speed}), together with the analysis in \cite{Dean}, we will find a stochastic partial differential equation for the dynamics of the coarse grained density field $\rho(\vec{r},t)$. We start by performing the second order Taylor expansion
\begin{widetext}
\begin{eqnarray}
\rho(\vec{r}, t+\Delta t) - \rho(\vec{r}, t) 
&=& \sum_{i=1}^N \left[ \delta(\vec{r}-\vec{r}_{i}(t+\Delta t)) - \delta(\vec{r}-\vec{r}_{i}(t)) \right]  \nonumber \\
&\approx& 	- \sum_{i=1}^N v_i(t) \, \hat{n}_{i}(t) \cdot \nabla \delta(\vec{r} - \vec{r}_{i}(t)) 
     		+ \frac{1}{2} \sum_{i=1}^N v_i^2(t) \, 
				{\hat{n}_i(t) \hat{n}_i(t) \colon \nabla \nabla \delta ( \vec{r} - \vec{r}_{i}(t) )}. 
\label{densityeq1}
\end{eqnarray}
\end{widetext}
Here, the operator `$\colon$' is the double-dot (or colon) product defined by 
$\vec{a} \, \vec{b} \, \colon \vec{c} \, \vec{d} = 
\sum_{\alpha} \sum_{\beta} a^{\alpha} b^{\beta} c^{\alpha} d^{\beta}$,
with indexes $\alpha$ and $\beta$ indicating the vector components.
By replacing $v_i(t)$ from the variable speed expressions (\ref{speed}) and (\ref{chi}), dividing by $\Delta t$
and taking the limit $\Delta t \rightarrow 0$, we find
\begin{widetext}
\begin{eqnarray}
\frac{\partial \rho}{\partial t} 
&\approx& -v_M    \sum_{i=1}^N 
	\bigg[ \frac{1}{{\mathcal{N}_{i}}^{\gamma}} |\sum_{j \in U_i} \hat{n}_{j}(t)|^{\gamma} \bigg] 
			\hat{n}_{i}(t) \cdot \nabla \delta(\vec{r} - \vec{r}_{i}(t)) \nonumber + 
	\frac{v_M^2}{2}    \sum_{i=1}^N 
	\bigg[ \frac{1}{{\mathcal{N}_{i}}^{2 \gamma}} |\sum_{j \in U_i} \hat{n}_{j}(t)|^{2\gamma} \bigg] 
			\hat{n}_{i}(t) \hat{n}_{i}(t) :\nabla \nabla \delta(\vec{r} - \vec{r}_{i}(t)) \nonumber \\ 
&=& 
-v_M \sum_{i=1}^N \frac{1}{{\mathcal{N}_{i}}^{\gamma/2}} 
	\bigg[1+ \frac{1}{\mathcal{N}_{i}}\sum_{j \ne j'} \hat{n}_{j} \cdot \hat{n}_{j'} \bigg]^{\gamma/2}
 		\hat{n}_{i}(t) \cdot \nabla \delta(\vec{r} - \vec{r}_{i}(t)) + \nonumber\\
&&  \; \; \;
\frac{v_M^2}{2} \sum_{i=1}^N \frac{1}{{\mathcal{N}_{i}}^{\gamma}} 
	\bigg[1 + \frac{1}{\mathcal{N}_{i}} \sum_{j \ne j'} \hat{n}_{j} \cdot \hat{n}_{j'} \bigg]^{\gamma}
 		\hat{n}_{i}(t) \, \hat{n}_{i}(t) : \nabla \nabla \delta(\vec{r} - \vec{r}_{i}(t)).
\end{eqnarray}
\end{widetext}
We can write this expression in terms of $\rho$, $\vec{P}$ and $\bf{Q}$ by using definitions (\ref{defrho}), (\ref{defp}) and (\ref{defq}). The resulting partial differential equation describes the dynamics of 
the density field in the Ginsburg-Landau approximation. It is given by
\begin{equation}
\frac{\partial \rho}{\partial t} = 
	- \frac{v_M}{\bar{m}^{\gamma/2}} \nabla \cdot (\vec{P} \rho)
	+ \frac{1}{2} \frac{v_M^2}{{\bar m}^{\gamma}}
						\nabla \nabla : ({\bf Q} + \frac{1}{2} {\bf 1}) \rho,
\label{rhovarspeed}
\end{equation}
where $\bar{m}(\vec{r},t)$ is a coarse-grained field representing the number of particles within a circle  of radius $\tilde r$ and centered at $\vec{r}$. More precisely, $\bar{m}(\vec{r}, t)= \pi \tilde r^2 \sum \rho(\vec{r}_j, t)$,
with the sum performed over all $j$ satisfying $|\vec{r}_j - \vec{r}| \le \tilde r$.

We are now in a position to compare Eq.~(\ref{rhovarspeed}) with the corresponding expression obtained in \cite{shradhasr} for particles moving with constant speed $v_C$:
\begin{equation}
\frac{\partial \rho}{\partial t} = -v_C \nabla \cdot (\vec{P} \rho)
+ \frac{1}{2} v_C^2\nabla \nabla : ({\bf Q} + \frac{1}{2} {\bf 1}) \, \rho .
\label{rhoconsspeed}
\end{equation}
Note that Eqs.~(\ref{rhovarspeed}) and (\ref{rhoconsspeed}) have the same form,
and become identical when we replace
\begin{equation}
v_C(\vec{r}, t) = \frac{v_M}{{\bar m(\vec{r}, t)}^{\gamma/2}}.
\label{varspeed}
\end{equation}
Hence, within the current approximation, we find that the particle speed and coarse-grained local number of neighbors will be correlated through an inverse power-law. This amounts to a relationship between local speed and local density that results from the imposed dependency of the particle speed on the local polarization order.
We confirm in Section \ref{Numerical} that this relationship is satisfied in numerical simulations, within the disordered regime.

The result in Eq.~(\ref{varspeed}), combined with the variable speed rule in Eqs.~(\ref{chi}) and (\ref{speed}), implies that regions of higher density will tend to be less ordered in this regime. This is in qualitative agreement with what we have observed experimentally, but is opposite to the typical relationship between {\it mean} density and {\it global} order found in minimal self-propelled particle models \cite{vicsekall,chateall}. 
Indeed, we will show numerically in Section \ref{Numerical} that, even in our current variable speed case, the level of polarization order of the whole system, $\psi$ (defined in Eq.~(\ref{eq:OrderParameter}) below), grows with its mean density $\rho_s$.
Despite its simplicity, our model is therefore able to capture a mechanism that relates local levels of density and order nontrivially if we impose the current particle speed rule. 

By following the same procedure as outlined above (see \cite{shradhasr} for details), we can also write equations of motion for the polarization order parameter. After a long but straightforward calculation, we find
\begin{equation}
\frac{\partial (\vec{P}\rho)}{\partial t} = F + G +  H, 
\label{pequation}
\end{equation}
where $F$ is the polynomial term, $G$ the derivative term, and $H$ is the noise term. 
The polynomial term is given by
\begin{equation}
F = \sqrt{\bar{m}} \left[ 1-2 \eta^2 - \frac{1}{\sqrt{\bar{m}}} - 
				\frac{1}{2} {\vec{P}} \cdot {\vec{P}} \right] {\vec{P}} .
\end{equation}
The derivative term is
\begin{widetext}
\begin{equation}
 G	=  -\frac{ v_M \sqrt{\bar{m}} }{ 2\bar{m}^{\gamma/2} } 
		\left[\vec{P}\nabla \cdot (\vec{P} \rho) + \nabla (|\vec{P}|^2 \rho) + 
			(\vec{P} \cdot \nabla)\vec{P} \rho  +  \nabla \cdot({\bf Q} \rho) + 
									\frac{1}{2}\nabla \rho \right]
		+    \frac{ v_M^2 \sqrt{\bar{m}} }{ 4\bar{m}^{\gamma} }
			\left[ \vec{T}+ \nabla^2(\vec{P}\rho) + \nabla(\nabla \cdot \vec{P}) \, \rho \right].
\label{derivativep}
\end{equation}
\end{widetext}
where we have simplified notation by introducing the vector $\vec{T}$, with 
$T_i =  \rho \, \nabla_{l} \nabla_{k}(Q_{lk} {P_i} + P_{l} P_{k} {P_i} + P_{l} Q_{ik})$
and indexes $i$, $k$, and $l$ labeling the corresponding tensor components.
Finally, the noise term (in the It$\hat{\rm{o}}$ interpretation) is found to be
\begin{equation}
H = \sqrt{\rho} \; {\bf M} \, \hat{h}.
\label{noisep}
\end{equation}
Here, $\hat{h}(\vec{r},t)$ is a vector field of unit length and random orientation, delta correlated in space 
and time, while ${\bf M}(\vec{r},t)$ is a $2 \times 2$ tensor field satisfying ${\bf M}^{2} = {\bf 1}$.

We now compare Eq.~(\ref{pequation}) to the constant speed case derived in \cite{shradhasr}. 
First, we find that the expressions for $F$ and $H$ remain unchanged. This implies that the transition point, which can be computed in both cases using a mean field approximation, will be the same. We confirm this result in Section \ref{Numerical} through numerical simulations.
By contrast, the derivative term (\ref{derivativep}) differs from the constant speed case, where we have
\begin{widetext}
\begin{equation}
 G =	  -\frac{ v_C \sqrt{\bar{m}} }{ 2 }
				\left[ \vec{P}\nabla \cdot (\vec{P} \rho) + \nabla (|\vec{P}|^2 \rho) + 
					(\vec{P} \cdot \nabla)\vec{P} \rho +  \nabla \cdot ({\bf Q} \rho) + 
											\frac{1}{2}\nabla \rho \right] + 
             		    \frac{ v_C^2 \sqrt{\bar{m}} }{ 4 } 
				\left[ \vec{T} + \nabla^2(\vec{P}\rho) + \nabla(\nabla \cdot \vec{P}) \, \rho \right].
\label{derivativepcons}
\end{equation}
\end{widetext}
Again, we find that the constant and variable speed cases are equivalent if we replace $v_C$ using Eq.~(\ref{varspeed}). 

We conclude that the hydrodynamic equations for the density and order parameter fields can be obtained in the variable speed case by simply replacing Eq.~(\ref{varspeed}) into the constant speed expressions.


\section{Numerical study \label{Numerical}}

In this section, we use numerical simulations to explore the range of validity of the analytic description derived above and study the dynamics of our variable speed model beyond this regime. 

We implemented agent-based simulations of $N = 2000$ particles in a two-dimensional periodic box of side $L$
using Eqs.~(\ref{position}), (\ref{velocity}) and (\ref{speed}). All runs presented in this Paper were carried out for an interaction range $\tilde r = 2.0$, maximum particle speed $v_M = 0.1$ and time-step $\Delta t = 1$, so that the particle displacement per time-step was never greater than $1/20$-th of the interaction radius. We studied simulations for different levels of noise $\eta$ and mean density $\rho_s$, using the variable speed rule in Eq.~(\ref{speed}) with exponent ranging from $\gamma = 0$ to $\gamma = 6$. The mean density was varied by changing the box size $L$ while keeping the total number of particles fixed.

We characterize the collective dynamics resulting from simulations using two different global order 
parameters. First, the global polarization order $\psi$, given by 
\begin{equation}
\psi(t) = \frac{1}{N} \sum_{i=1}^{N} \hat{n}_i(t).
\label{eq:OrderParameter}
\end{equation}
This measures the degree of alignment in the system. It is equal to $1$ when all particles are heading in the same direction, and to $0$ when they are randomly oriented, regardless of the particle speeds.
Second, the mean particle speed $\bar{v}$, defined as
\begin{equation}
\bar{v}(t) = \frac{1}{N} \sum_{i=1}^{N} v_{i}(t).
\label{eq:MeanSpeed}
\end{equation}
This order parameter is equal to $0$ when all particles are immobile and to $v_M$ when they advance at their maximal speed. Note that both quantities are defined here for a specific simulation snapshot at time $t$. We will use below these instantaneous values and their averages over time: $\left< \psi \right>$ and $\left< \bar{v} \right>$.

\subsection{Range of validity of analytical results}

In order to verify the validity of the analytical derivations in Section \ref{Analytical}, we confirmed that the relation in Eq.~(\ref{varspeed}) holds approximatively for a range of simulation parameters. We performed runs using $\gamma = 6$, $ \rho_s = 2.5$ and various levels of noise $\eta$. We then measured the speed $v_i$ of each particle and the number of neighbors $\mathcal{N}_{i}$ within its interaction range (including itself). 

\begin{figure}
\begin{center}
{\includegraphics[width=8.6cm]{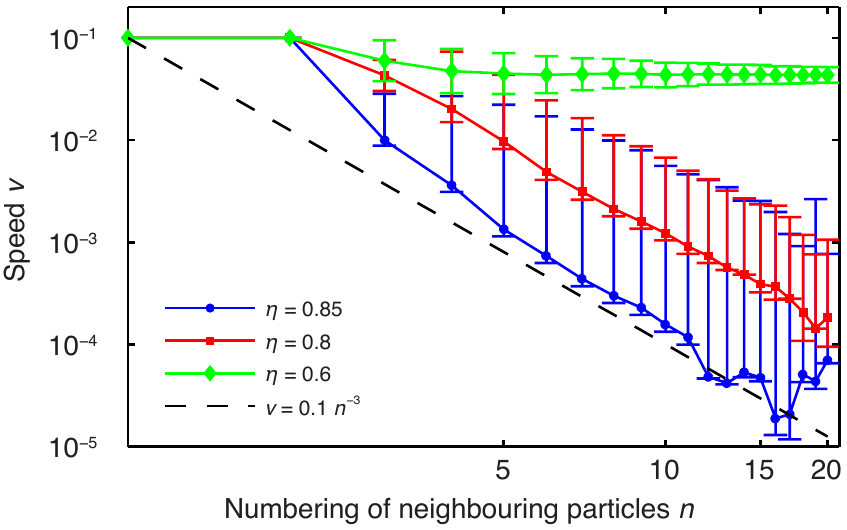}}
\caption{Relationship between the particle speed and local density that emerges from simulations of our minimal 
variable speed model for $\gamma = 6$. The top and bottom error bars show the standard deviations computed using only speeds above or below the mean, respectively.
The dashed line displays the $v = v_M \mathcal{N}^{\gamma/2}$ relationship deduced analytically in the continuous approximation. As the noise is increased and local ordered structures vanish, this approximation becomes more valid.
}
\label{fig02}
\end{center}
\end{figure}

The average speed of all particles with a given number of neighbors $\mathcal{N}$ is plotted on Fig.~\ref{fig02}. In regimes where our analytical approach is valid, Eq.~(\ref{varspeed}) implies that this speed should be approximately equal to $v_M / \mathcal{N}^{\gamma/2}$.
The figure shows that this relationship is satisfied for high levels of disorder, where no ordered structures that could violate the approximations in our Ginzburg-Landau approach are present. We find that for $\eta \geq 0.7$ the inverse power-law proportionality is already verified and that above $\eta \approx 0.85$ the exact analytical dependency in Eq.~(\ref{varspeed}) is approximately followed (displayed on the figure as a dashed line). The analytic approximation fails, however, for noise values $< 0.7$, where particles are more ordered. 

\subsection{The order-disorder transition}

It is common to find in self-propelled particle models, such as the Vicsek model \cite{vicsek}, a non-equilibrium phase transition that separates the disordered state where particles move in random directions from the ordered one where they have a common heading. The ordered phase can be reached by decreasing the noise level or by increasing the mean density \cite{chatePRL,chate2,chuepe1,chuepe2,chuepe3}. This transition is also present in a variation of the Vicsek model introduced by Gr\'egoire and Chat\'e in \cite{chatePRL} that is almost identical to our current model, but with constant particle speed (i.e., our $\gamma=0$ case). We will examine below the effect of the variable speed rule on this transition.

\begin{figure}
\begin{center}
{\includegraphics[width=86mm]{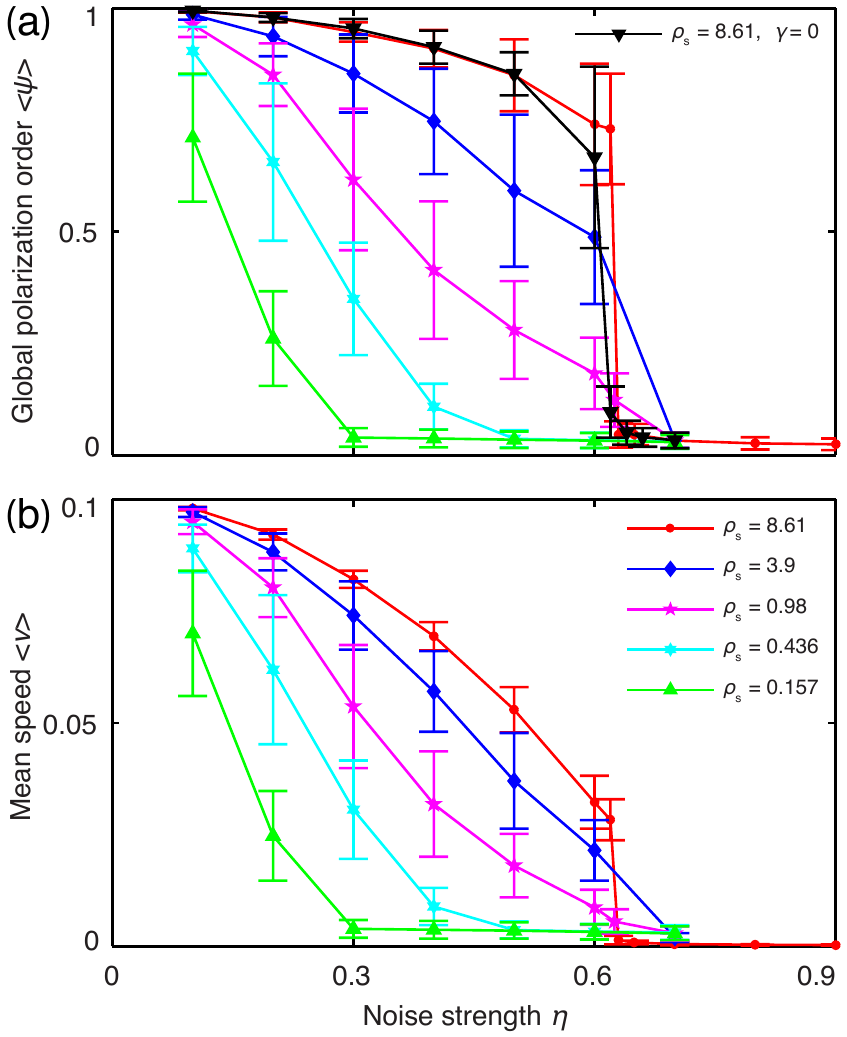}}
\caption{Mean global polarization order $\left< \psi \right>$ (a) and mean speed $\left< v \right>$ (b) as a function of the noise strength
$\eta$ for different values of the mean density $\rho_s$. 
At $\eta_c \approx 0.63$, the system undergoes a discontinuous transition from an ordered to a disordered state.
This value appears unchanged for the $\gamma = 0$ case (constant speed) and for $\rho_s \ge 0.98$. 
As the system approaches the disordered state, $\left< v \right>$ decreases, vanishing for $\eta > \eta_c$. All curves were computed using $N=2000$ particles, interaction radius $\tilde r = 2.0$, maximum particle speed $v_M = 0.1$, and variable speed parameter $\gamma = 6.0$. The different densities were obtained by changing the system size, with $L = 27$, $40$, $80$, $120$, and $200$.}
\label{fig03}
\end{center}
\end{figure}
\begin{figure}
\begin{center}
{\includegraphics[width=86mm]{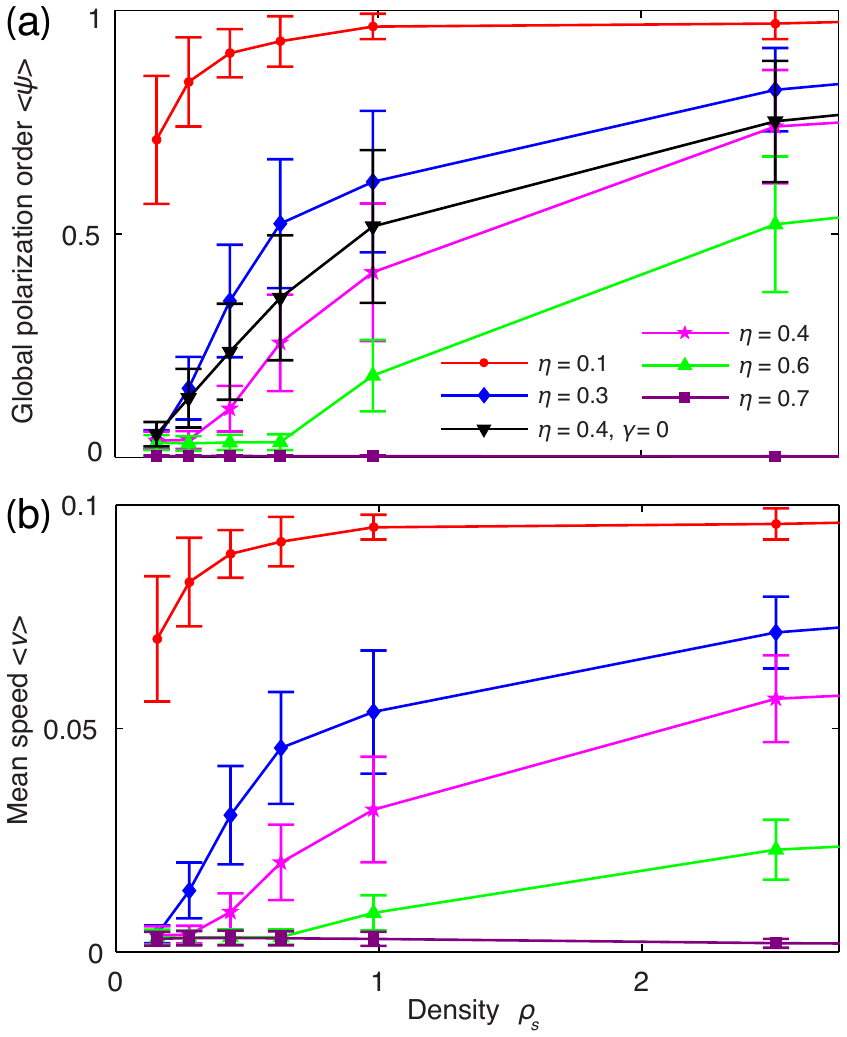}}
\caption{Mean global polarization order $\left< \psi \right>$ (a) and mean speed $\left< v \right>$ (b) as a function of the 
mean density $\rho_s$ for different levels of noise $\eta$. All other parameters are the same as in
Fig.~\ref{fig03}. As $\rho_s$ is decreased, the system undergoes a transition 
from an ordered state with high $\left< \psi \right>$ and $\left< v \right>$ values to a disordered state with 
$\left< \psi \right> \approx \left< v \right> \approx 0$. The transition displayed by the order parameter $\left< \psi \right>$ is very 
similar for the constant speed case displayed ($\gamma = 0$). For $\eta = 0.7 > \eta_c$ no transition is 
observed, since the stationary solutions remain disordered for all values of the density $\rho_s$.}
\label{fig04}
\end{center}
\end{figure}

Figure \ref{fig03} displays the mean polarization order parameter $\left< \psi \right>$ and particle speed $\left< v \right>$ as functions of the noise $\eta$ for different values of $\rho_s$. We observe that for values of $\rho_s \ge \sim 1$ there is a sharp discontinuous transition point at $\eta_c \approx 0.6$. For lower values of $\rho_s$, the transition appears smoother and, as expected, occurs at lower critical noise levels. The decrease in the order parameter $\left< \psi \right>$ is accompanied by a reduction of the mean particle speed. Note that $\left< v \right>$ is already substantially reduced for $\eta$ values where $\left< \psi\right>$ is still high.
In addition to the curves obtained with the variable speed model, we also include in Fig.~\ref{fig03}(a) a curve displaying the transition for a constant speed case ($\gamma = 0$) with $\rho_s = 8.61$.
We observe that the transition point does not change strongly, as predicted by the analytical results
presented in Section \ref{Analytical}. We verified that this is also the case for other values
of $\gamma$ between $0$ and $6$ (data not shown). 

Figure \ref{fig04} shows the same order-disorder transition as Fig.~\ref{fig03}, but now as a function of the mean density $\rho_s$, for different values of $\eta$. We find again that the mean particle speed decreases as the system loses order. For $\eta > 0.6$, the dynamics remain disordered even at high density values, which is consistent with the results in Fig.~\ref{fig03}. As the noise level is reduced, the critical $\rho_s$ decreases until it reaches a point where no transition is observed even for very low mean denisity values. Finally, we include again in Fig.~\ref{fig04}(a) a curve for $\gamma = 0$ to show that the transition point also remains unchanged here with respect to the constant speed case.

We conclude from this analysis that the main features of the order-disorder transition remain unchanged in the variable speed case, as predicted by our analytical calculations. However, the critical slowdown of particles that is shown above to occur as the system loses order has significant effects in the resulting collective dynamics. We will study this phenomenon in more detail below.

\subsection{Bi-stable solutions}

We now take a closer look at the order-disorder transition. Our aim here is not to extrapolate its properties to the thermodynamic limit, as this would require larger computations and a systematic finite size scaling analysis. Instead, we focus on understanding how the variable speed affects the particle dynamics near the transition.

\begin{figure*}
\begin{center}
{\includegraphics[width=17.8cm]{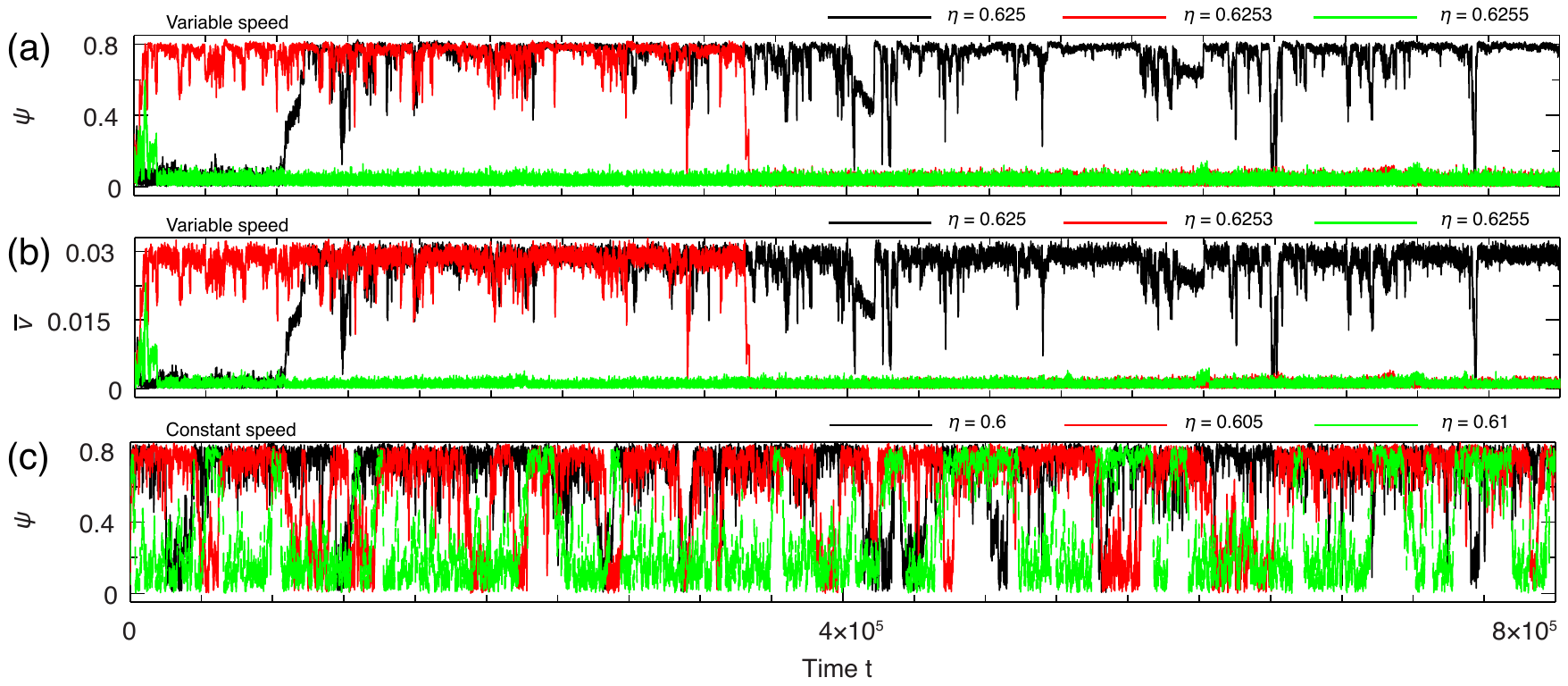}}
\caption{Global polarization order $\psi(t)$ and mean speed $\bar{v}(t)$ per frame, as a function of time. 
The variable speed (a and b) and constant speed (c) cases are displayed for the same parameters used in
Fig.~\ref{fig03}, with $\rho_s = 8.61$ and different noise strengths, close to the 
transition point. Bistable solutions are observed in both cases, but in the variable speed case the transition to 
the disordered branch is accompanied by a critical slowdown of the dynamics, due to the imposed relationship
between particle speed and local polarization order. The intermittent switching between states thus becomes much less frequent. Note that the critical noise value $\eta_c$ is slightly higher in the variable speed case.}
\label{fig05}
\end{center}
\end{figure*}
\begin{figure}
\begin{center}
{\includegraphics[width=8.6cm]{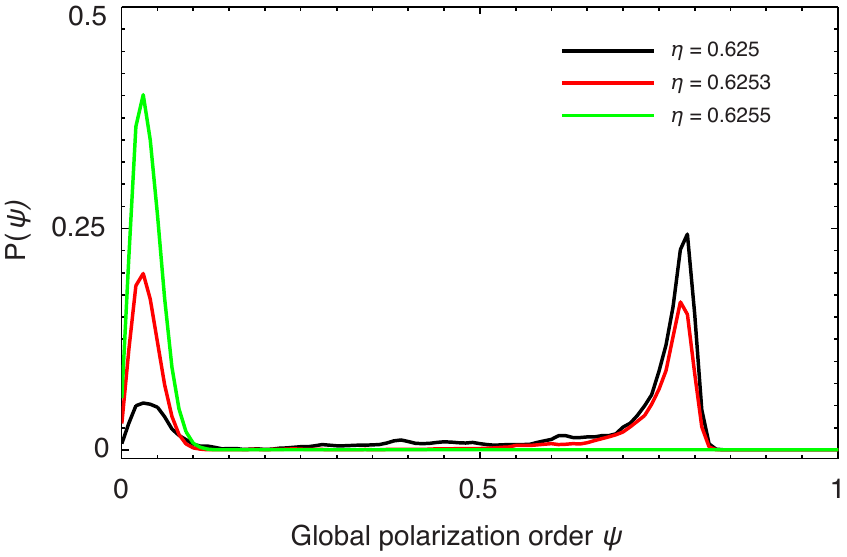}}
\caption{Distribution of the global order parameter $\psi$ for the variable speed case displayed in 
Fig.~\ref{fig05}(a).
The system exhibits a discontinuous transition with two coexisting states near the critical point.}
\label{fig06}
\end{center}
\end{figure}

Figure \ref{fig05}(a) displays the global order parameter $\psi$ as a function of time for three different values of $\eta$ close to the critical noise. The corresponding histograms are shown in Fig.~\ref{fig06}.
We observe that the dynamics is bistable near the transition; the system switches between ordered and
disordered states. This is in agreement with the first-order transition that had been previously reported in the constant speed case, both in simulations and in the mean-field approximation \cite{chatePRL,chate2,chuepe1,chuepe2,chuepe3}.
We confirm in our simulations that the $\gamma = 0$ case also displays bistable dynamics near the critical point, as shown in Fig. \ref{fig05}(c). We find, however, that there is a strong difference between the time series obtained in the variable speed and constant speed cases. While the former presents very few transitions between states, even for the long time-series presented, the latter shows several intermittent jumps from one state to the other \cite{chuepe4}.
This difference is explained by the lower speed imposed in our model to particles surrounded by low local order, which results in a critical slowdown of the global dynamics as the order parameter $\psi$ fluctuates to its 
lower metastable state. Indeed, Fig.~\ref{fig05}(b) displays the mean particle speed $\bar{v}(t)$ for the same simulation used to generate Fig.~\ref{fig05}(a). The corresponding curves are strongly correlated, with the average global dynamics following the local variable speed rule. 
The disordered branch of the bistable solution therefore acts similar to an absorbing state; once it is reached the particle dynamics slow down so much that it is very hard to escape.

\subsection{Spatial dynamics and phase segregation}

We turn our attention to the spatial dynamics observed in the variable speed model. 
This is where the strongest differences with the constant speed case emerge. 
We find that, in the disordered state close to the transition, a phase separation occurs where some 
particles condense into almost immobile high-density clusters while the rest form a low-density ballistic 
gas between them.

\begin{figure*}
{\includegraphics[width=17.8cm]{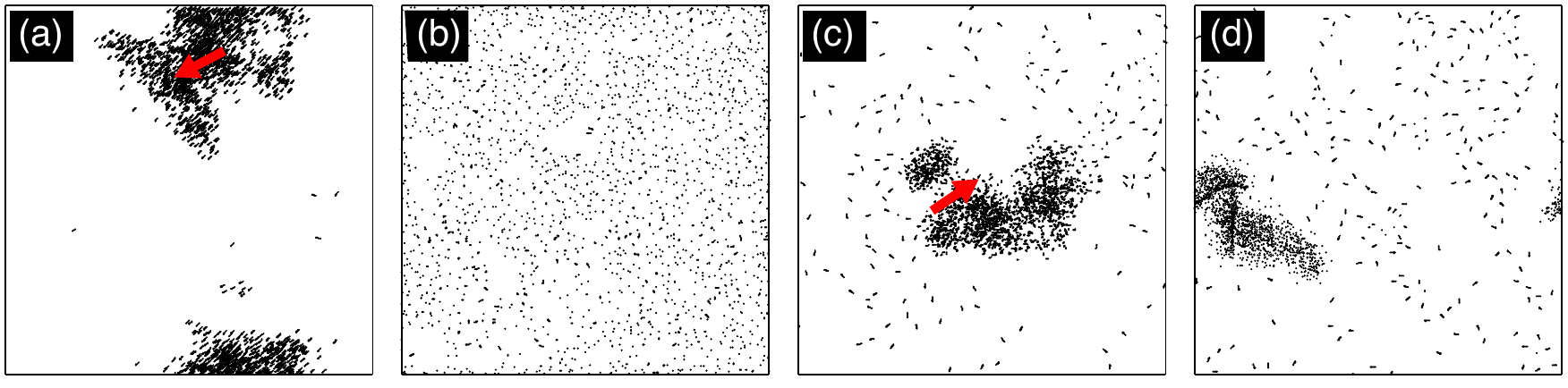}}
\caption{Simulation snapshots for $N=2000$ particles in four different steady states with the 
same mean density $\rho_s = 8.61$, maximum particle speed $v_M = 0.1$, and variable speed exponent 
$\gamma = 6$. Each particle is represented by a short line with length proportional to its speed. 
(a) Low noise ($\eta = 0.1$) ordered state. We observe clusters of moving particles (with velocity indicated by 
the arrow) equivalent to those found in models with constant speed.
(b) High noise ($\eta = 0.7$) disordered state. Particles are almost immobile and homogeneously distributed in 
space. Due to the high $\rho_s$ value considered here, each particle interacts with several neighbors heading in different directions, resulting in low particle speed.
(c and d) Intermediate noise states ($\eta = 0.6253$) close to the transition. The system here can be either in the
ordered solution branch (c), displaying a moving cluster (with velocity indicated by the arrow), or in the disordered
branch (d), which exhibits a static cluster as described in the main text.}
\label{fig07}
\end{figure*}

Figure \ref{fig07} presents simulation snapshots of the variable speed model with $\gamma = 6$, $N=2000$, $\rho_s = 8.61$ and three different levels of noise. The length of each line is proportional to the speed of the corresponding particle. The four panels display: an ordered state (a), a disordered state (b), and the two branches of the bistable state discussed in the previous section (c and d).
Snapshot (a) corresponds to a simulation with low noise $\eta = 0.1$. The system is ordered ($\eta$ is much lower than its critical value $\eta_c \approx 0.625$), with particles moving in the same approximate direction. Since local neighborhoods are mostly ordered, $\chi$ takes values close to $1$ in Eq.~(\ref{speed}) and particles advance at speeds close to $v_M$.
By contrast, snapshot (b) corresponds to a high noise, $\eta = 0.7$, disordered case. Here, particles are almost immobile. Due to the high mean density, even if the noise spreads out particles almost homogeneously, each one has several neighbors within its interaction range. Given the low local order, $\chi \approx 0$ in Eq.~(\ref{speed}) and particle speeds are close to zero. 

Snapshots (c) and (d) are particularly interesting as they display previously unobserved dynamics, not possible in a constant speed model, that occur when the noise is close to its critical value. As shown in Figs.~\ref{fig05} and \ref{fig06}, we have in this case bistable global dynamics. The two metastable solution branches correspond to an ordered state with high $\psi$ and to a disordered state with low $\psi$. 
Both states organize into a high-density cluster surrounded by a low-density particle gas. In panel (c), the cluster moves as indicated by the red arrow, while in panel (d) it remains almost static. Larger simulations can display several clusters with similar behavior. Particles surrounding the clusters move rapidly. 

While the presence of moving clusters is common in Vicsek-like models \cite{chuepe4,chuepe5}, static clusters such as the one on Fig.~\ref{fig07}(d) have not been previously observed. The mechanism that leads to their nucleation therefore deserves more a detailed analysis. As the system fluctuates between an ordered and a disordered state (being close to the transition point), it reaches situations where particles are typically not locally aligned. This leads to the formation of groups of slow moving particles that grow as other particles reach them, because incoming particles suddenly confront high density disordered regions, thereby losing orientation and slowing down drastically. Once clusters are formed, they can only lose particles at their frontier. There, some particles will spontaneously align due to noise fluctuations and manage to escape if they head away from the cluster, since they will feel less and less the influence from the disordered region. Eventually a single particle, or a small group, will be far enough removed to feel only its own influence, forming a low density ballistic gas between clusters. Particles in this gas are isolated and their $\chi$ is therefore close to $1$. They move at speeds close to $v_M$ until they reach another cluster and condense again. These clusters will thus grow until their absorption rate is balanced by their evaporation rate.
In the bulk of the clusters particles advance very slowly, which hinders their ability to reorder. Indeed, it has been shown that, when noise is present, Vicsek-like models can only reach an ordered state if particles are able to move with respect to each other, which allows them to switch neighbors and establish effective long-range interactions over time \cite{chuepe3}.
If particles always interact with the same neighbors, the system becomes equivalent to an XY-model, for which the Mermin-Wagner theorem shows that no long-range order can exist in two dimensions at nonzero noise levels \cite{merminwagner}.
Hence, the lack of relative particle motion in the bulk of a static cluster helps stabilize it in a disordered, immobile state.

\begin{figure}
\begin{center}
{\includegraphics[width=86mm]{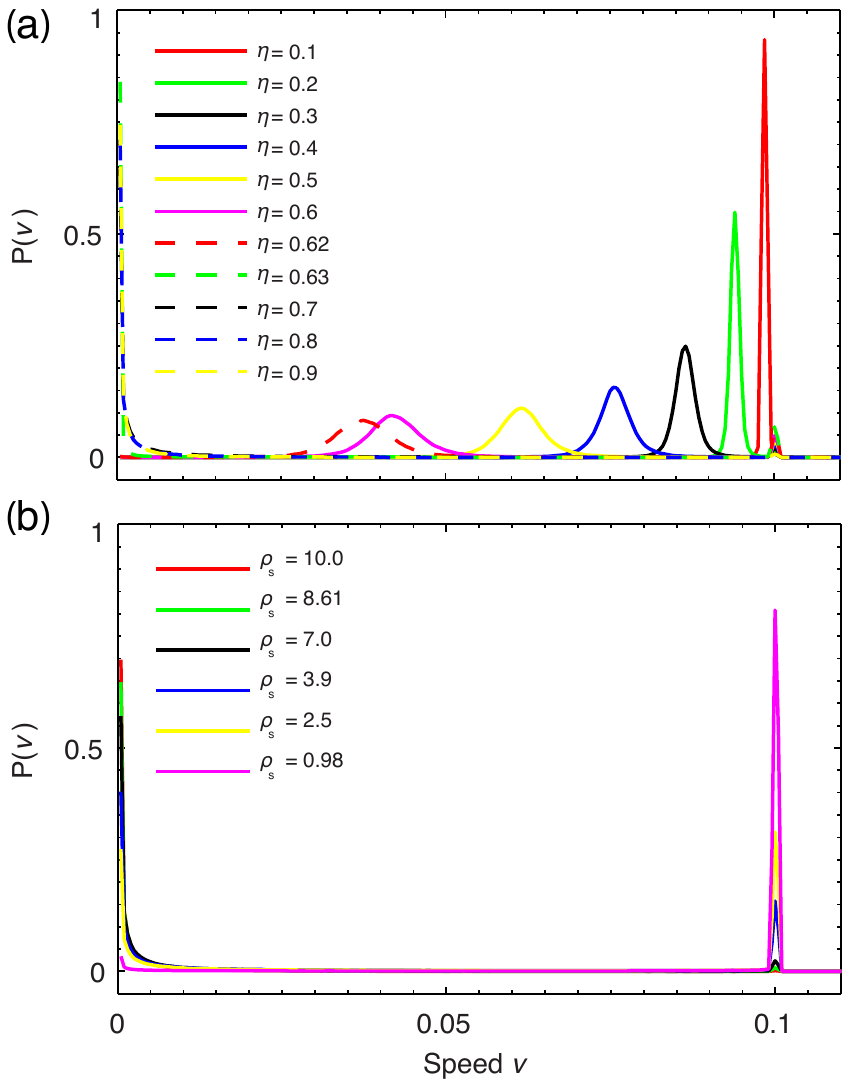}}
\caption{
(a) Distribution of individual particle speeds for the same parameters as in Fig.~\ref{fig03}, with $\rho_s = 8.61$ and different levels of noise intensity. As the noise level is increased, the typical particle speed decreases until static clusters start nucleating for $\eta \ge \eta_c \approx 0.625$. These clusters absorb most particles, while a few isolated particles continue to move between them at maximal speed, as shown by the peaks at $v = 0$ and $v = v_M = 0.1$.
(b) Distribution of individual particle speeds for the same parameters as in Fig.~\ref{fig04}, with $\eta = 0.7$ (disordered state) and different values of the mean density $\rho_s$. Here, static clusters are always present, their size increasing with $\rho_s$. Particles are again either frozen within these clusters or moving alone between them at maximal speed.}
\label{fig08}
\end{center}
\end{figure}

\begin{figure}
\begin{center}
{\includegraphics[width=86mm]{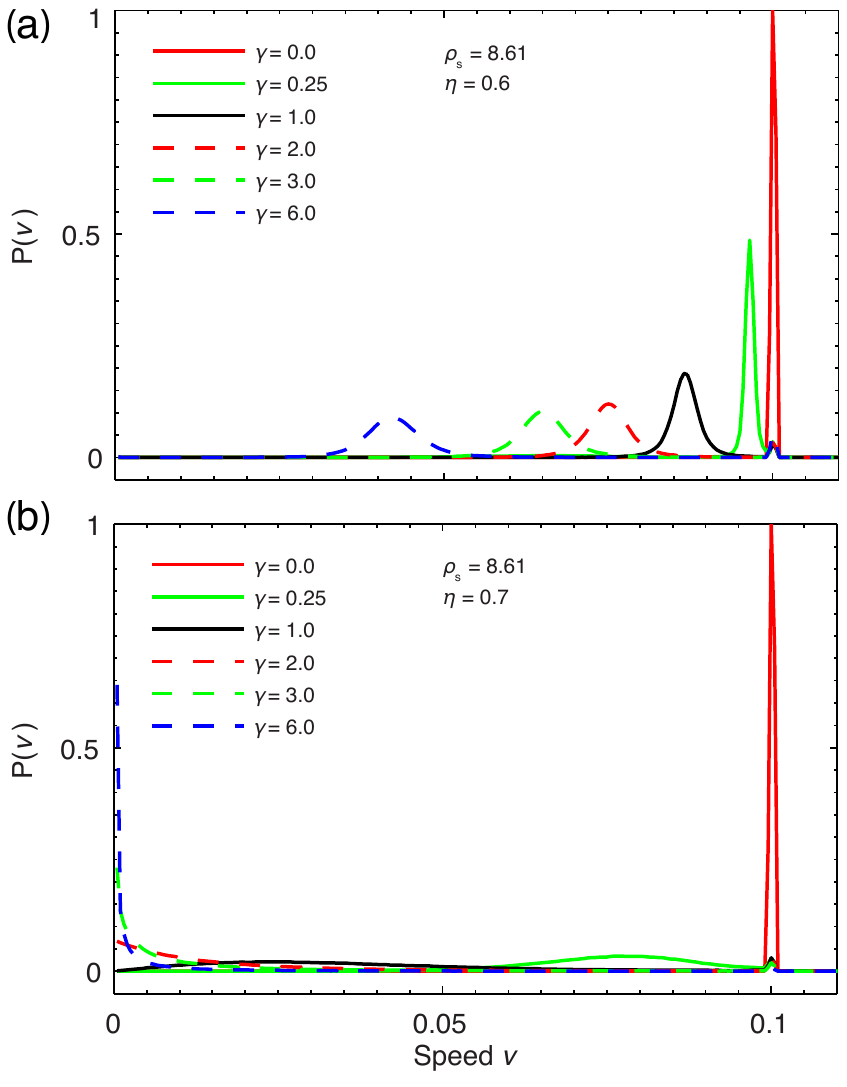}}
\caption{
Distribution of individual particle speeds for the same parameters as in Fig.~\ref{fig03}, with noise intensity just below ($\eta~=~0.6$) or above ($\eta= 0.7$) the critical noise value, and different values of the variable speed exponent $\gamma$.
When in the ordered state (a), the typical particle speed gradually decreases as $\gamma$ is increased, from a fixed 
$v = v_M = 0.1$ value for $\gamma = 0$, to $v \approx 0.04$ for $\gamma = 6$.
In the disordered state (b), by contrast, the typical particle speed is reduced much more abruptly for any 
$\gamma > 0$, where the static clusters described in the main text are nucleated.}
\label{fig09}
\end{center}
\end{figure}

In order to study how the particles distribute between those that are part of a static cluster and those that are not, we plot on Fig.~\ref{fig08} the individual particle speed distribution for various values of noise and density.
Panel (a) shows that, as the noise level is incresased, first the typical particle speed is smoothly reduced and then a zero-speed peak emerges, corresponding to static clusters. While in the ordered phase, groups of particles therefore move slower and slower until the critical noise level is reached and static clusters start nucleating.
In panel (b) the noise level is fixed at a high value ($\eta = 0.7$), so the system is always in the disordered state. As we vary the density $\rho_s$, the particle speed distribution remains bimodal. Particles can either be within a static cluster and stop advancing or be part of the inter-cluster gas and move at maximum speed $v_M$ while remaining isolated. As the mean density is increased, more particles are trapped in static clusters and thus have zero speed.

Finallly, we present in Fig.~\ref{fig09} the individual particle-speed distribution as a function of the variable speed exponent $\gamma$. We consider two parameter regimes, one with noise intensity just below the critical noise value [Fig.~\ref{fig09}(a), $\eta = 0.6$] and one just above it [Fig.~\ref{fig09}(b), $\eta = 0.7$].
In the first situation, the typical particle speed gradually decreases as $\gamma$ is increased, from $v \approx v_M = 0.1$ when $\gamma = 0$, to $v \approx 0.04$ for $\gamma = 6$. 
In the disordered state, by contrast, the typical particle speed is reduced much more abruptly for any $\gamma > 0$, quickly giving rise to the nucleation of static clusters.


\section{Discussion \label{Discussion}}

The analysis above demonstrates the similarities and differences between standard, minimal constant speed models of collective motion \cite{vicsek,chateall} and our variable speed version. We show that the variable speed case displays an order-disorder transition analogous to that observed for constant speed. In both cases, as the noise level is decreased or the mean density increased, the system goes from a disordered state to an ordered one where they align to a common heading. The dynamics associated to the transition, however, are very different.

An interesting finding is that our variable speed rule induces an inverse power-law relationship between the speed of a particle, or the level of order in its surroundings, and the local density.
Note that this is opposite to the typical correlation between {\it global} order and {\it mean} density in minimal models \cite{vicsek}. Indeed, high mean density makes particles interact in average with more neighbors, which increases the level of order, as it has been shown numerically and analytically (in the mean field approximation) \cite{chuepe3,chuepe2}.

While the model presented in this paper is far from realistic, the inverse relationship between local quantities specified above appears to be robust enough to persist in less idealized systems.
In the golden shiner data we find not only the correlation between individual speed and local polarization order that inspired our variable speed rule (see Fig.~\ref{fig01}), but also an inverse relationship between local order, or particle speed, and local density analogous to that deduced from our model (data not shown). We emphasize that we cannot deduce  a causal relation between these effects from the experimental data. By establishing connections between them in a specific model, however, we have provided a mechanism through which they could be robustly related in generic situations.

Another observation from our study is the nucleation of static clusters. In simulation videos the dynamical process leading to these clusters looks similar to jamming in granular materials \cite{JammingRefs}. Despite the differences between these processes, we can draw some analogies that go beyond their visual appearance. In physical systems, the opposing forces on jammed particles add to zero, stopping their flow. Likewise, here opposing headings (i.e. conflicting information) add up to zero, producing no local order and therefore a vanishing particle speed.
We also note that both jammed and static regions will grow by recruiting moving particles that reach them and are stopped by opposing interactions.
Finally, an additional similarity is given by the dynamics in the bulk and edges of the clusters. In both cases edge particles can escape if the sum of all surrounding interactions points away from the cluster, while bulk particles can only move if there is a pathway of particles with a non-zero sum of interactions along it that percolates through the group \cite{JammingRefs}.

Despite the similarities outlined above, the mechanisms behind both processes are very different. Jamming is produced by contact forces or repulsive potentials, while interactions in the variable speed model are based on heading directions. Another important difference is that in our model static clusters cannot form below a certain noise level, while jamming is always increased at lower noise levels.

We end this section by pointing out that our variable speed model considers only one possible way of relating the speed of an individual particle to its local environment. While the model was inspired by experimental data, we do not claim any specific causal origin to this correlation. It could result from the interplay of a number of biological interactions that are not considered in this minimal model. It could also be for example, that alignment is enhanced in faster moving particles or that the speeding rule depends in fact directly on the local density.


\section{Conclusion \label{Conclusion}}

We have studied the dynamics of a minimal model of collective motion where the particles move with variable speed. We found that, despite the simplicity of the algorithm and its similarity with standard constant speed models, our system displays not only the usual ordering transition but also new generic dynamical phenomena that could be present in more realistic models or in experimental systems. These are: (1) an inverse correlation between individual particle speed or local polarization order and local density, produced by the imposed relationship between particle speed and local order, and (2) the nucleation  in certain regimes of static clusters where individuals remain immobile, only turning in place without achieving order.

We conclude that including variable speed dynamics in standard self-propelled particle models produces a range of new phenomena that can be relevant for experimental systems. Given the amount of work dedicated up to now to models with constant speed, we hope that our work will serve as a motivation to explore the generic and specific consequences of considering variable speed rules in this class of systems.


\begin{acknowledgments}
This work was partially supported by the National Science Foundation under 
Grant No.~PHY-0848755. 
SM would like to thank Prof. Sriram Ramaswamy for his help in the derivation of 
hydrodynamic equations for the constant speed model. KT acknowledges support from the Research Council of Norway.
\end{acknowledgments}



\end{document}